\documentclass[sigconf]{acmart}
\AtBeginDocument{%
  \providecommand\BibTeX{{%
    \normalfont B\kern-0.5em{\scshape i\kern-0.25em b}\kern-0.8em\TeX}}}

\copyrightyear{2021}
\acmYear{2021}
\setcopyright{acmlicensed}
\acmConference[AISec '21]{Proceedings of the 14th
ACM Workshop on Artificial Intelligence and Security}{November 15,
2021}{Virtual Event, Republic of Korea}
\acmBooktitle{Proceedings of the 14th ACM Workshop on Artificial Intelligence
and Security (AISec '21), November 15, 2021, Virtual Event, Republic of Korea}
\acmPrice{15.00}
\acmDOI{10.1145/3474369.3486877}
\acmISBN{978-1-4503-8657-9/21/11}

\usepackage{xcolor,colortbl}
\usepackage{tikz}
\newcommand{\SG}[1]{(#1)}
\newcommand{\AC}[1]{A#1.}
\newcommand{\crow}{\cellcolor{blue!5}}
\begin{document}
\fancyhead{}
\title[Automating Privilege Escalation with Deep Reinforcement Learning]{Automating Privilege Escalation with\\ Deep Reinforcement Learning}

\author{Kalle Kujanp{\"a}{\"a}}
\email{kalle.kujanpaa@aalto.fi}
\affiliation{%
  \institution{Aalto University}
  \city{Espoo}
  \country{Finland}}

\author{Willie Victor}
\email{willie.victor@f-secure.com}
\affiliation{%
  \institution{F-Secure}
  \city{Johannesburg}
  \country{South Africa}}

\author{Alexander Ilin}
\email{alexander.ilin@aalto.fi}
\affiliation{%
  \institution{Aalto University}
  \city{Espoo}
  \country{Finland}}


\begin{abstract}
AI-based defensive solutions are necessary to defend networks and information assets against intelligent automated attacks. Gathering enough realistic data for training machine learning-based defenses is a significant practical challenge. An intelligent red teaming agent capable of performing realistic attacks can alleviate this problem. However, there is little scientific evidence demonstrating the feasibility of fully automated attacks using machine learning. In this work, we exemplify the potential threat of malicious actors using deep reinforcement learning to train automated agents. We present an agent that uses a state-of-the-art reinforcement learning algorithm to perform local privilege escalation. Our results show that the autonomous agent can escalate privileges in a Windows~7 environment using a wide variety of different techniques depending on the environment configuration it encounters. Hence, our agent is usable for generating realistic attack sensor data for training and evaluating intrusion detection systems.
\end{abstract}




\maketitle

\section{Introduction}

Defending networks and information assets from attack in a constantly evolving threat landscape remains a substantial challenge in our modern and connected world. As better detection and response methods are developed, attackers invariably adapt their tools and techniques to remain competitive. One adaptation is the use of advanced automation to perform attack sequences so quickly they cannot be responded to by defenders. An example of this is the NotPetya malware from 2017. The malware spread very rapidly by credential dumping and lateral movement techniques usually expected of human-on-keyboard attacks (see e.g. \cite{greenberg2018untold}).

Partial or even full automation of attacks is nothing new. Many exploits and techniques consist of discrete steps, and they can be easily scripted. Frameworks such as Metasploit, OpenVAS, Cobalt Strike, and PowerShell Empire support and automate red teaming activities. However, this type of automation relies on several assumptions about the target environment and often requires a human to configure it for a specific scenario. Moreover, the highly predictable sequences of observable events generated by automated attacks give a further advantage to defenders. Hence, these kinds of attacks can often be detected and responded to efficiently. In contrast, a human expert might be capable of determining the optimal approach for a given target almost immediately and avoid detection, thanks to the experience that they have gathered.

Can the human approach be emulated by intelligent machine learning agents that learn from experience, instead of the user having to enumerate all possibilities and creating logic trees to account for all the alternatives? There would be many potential use cases for an intelligent machine learning red teaming agent. To train comprehensive defensive systems using machine learning, enormous amounts of data of realistic attacks might be necessary. An intelligent agent enables the generation of large amounts of attack data, on-demand, that could be used to train or refine detection models. Moreover, in order to understand how to defend against attacks and perform risk assessment, the potential behavior and threats posed by these machine learning-based agents must be understood by the cyber security community. 

Despite a fear of malicious actors using reinforcement learning for offensive purposes and the significant advances in deep reinforcement learning during the past few years, few studies on red teaming with reinforcement learning have been published.
The most likely reason for this is that the problem of learning to perform an attack is extremely hard:
\begin{itemize}
    \item a complete attack typically consists of a long sequence of interdependent steps;
    \item the action space is practically infinite if the actions are the commands that the agent can execute;
    \item even formalizing red teaming activities as machine learning problems can be extremely challenging. 
\end{itemize}
Therefore, existing research has focused on automating smaller sub-tasks such as initial access \cite{deepexploit} or lateral movement during post-exploitation \cite{maeda2021automating}. 

DeepExploit \cite{deepexploit} is a deep reinforcement learning agent that is trained to automate gaining initial access using known vulnerabilities and exploits. It is built on the Metasploit framework \cite{metasploit}. After a successful penetration, it tries to recursively gain access to other hosts in the local network of the given input IP address.
DeepExploit is primarily a framework for penetration testing, and its support for post-exploitation activities is very limited: the agent treats lateral movement as a second initial access task.

The deep RL agent of Maeda and Mimura \cite{maeda2021automating} is a step forward in emulating adversarial behavior in real environments: it is trained to perform lateral movement in Windows domains. The authors train the agent using the modules of PowerShell Empire as the action space. The state of the proposed agent consists of ten entries, such as the number of discovered computers in the network, the number of compromised computers, and whether the agent has local administrative privileges. The authors demonstrate that the reinforcement learning agent can learn to perform lateral movement and obtain domain controller privileges.

In this work, we present one potential use case for an intelligent machine learning red teaming agent: we use deep RL to automate the task of local privilege escalation. Privilege escalation is the typical first step performed by an attacker after gaining initial access, and it is often followed by lateral movement to other hosts in the penetrated network. We consider privilege escalation in Windows~7 environments, which may have an arbitrary number of system components (such as services, DLLs, and tasks). We propose a formalization of the privilege escalation task as a reinforcement learning problem and present a novel architecture of an actor-critic RL agent. We experimentally show that the proposed agent can learn to perform the privilege escalation task.

Although we focus on one sub-task performed by a malicious actor,
the learning problem that we consider is significantly harder compared to previous works \cite{deepexploit,maeda2021automating}:
\begin{itemize}
\item Privilege escalation needs a sequence of actions, the selection of which depends on the changing system state. The scenario of \cite{deepexploit}, for example, has one-step solutions without any changes in the system state.

\item Privilege escalation can be accomplished by multiple different strategies.

\item The attacked system can have a varying number of system components (services, DLLs, tasks), and the agent should generalize to any number of those.

\item Our training setup is more realistic and diverse compared to \cite{maeda2021automating}. Instead of attacking a system whose variability is implemented by noise in system parameters, we use different system configurations in each training episode.

\item The state of the system in our experiments is described by thousands of variables, which is much larger than the states of the agents in \cite{maeda2021automating,deepexploit}.

\item The actions that we design for our learning environment are more atomic compared to \cite{deepexploit,maeda2021automating}. Most of our actions can be implemented with standard OS commands instead of using modules of existing exploitation frameworks.
\end{itemize}
Thus, our study takes a step towards solving more complex red teaming tasks with artificial intelligence.

\section{Related Work}

Applying reinforcement learning in cyber security has been a subject of much recent research \cite{nguyen2019deep}. Examples of application areas include, among others, anti-jamming communication systems \cite{han2017two}, spoofing detection in wireless networks \cite{xiao2015spoofing}, phishing detection \cite{chatterjee2019detecting}, autonomous cyber defense in software-defined networking \cite{han2018reinforcement}, mobile cloud offloading for malware detection \cite{wan2017reinforcement}, botnet detection \cite{alauthman2020efficient}, security in mobile edge caching \cite{xiao2018security}, and security in autonomous vehicle systems \cite{ferdowsi2018robust}. In addition, reinforcement learning has been applied to research of physical security, such as grid security \cite{ni2019multistage}, and to green security games \cite{wang2019deep}.

Previously, multi-agent reinforcement learning has been applied to cyber security simulations with competing adversarial and defensive agents \citep{bland2020machine, elderman2017adversarial, he2016faster}. It has been shown that both the attacking and the defending reinforcement learning agents can learn to improve their performance. The success of multi-agent reinforcement learning might have wider implications for information security research even though these simulation-based studies are not directly applicable to real environments.

There have also been attempts to apply reinforcement learning to penetration testing \citep{deepexploit, ghanem2018reinforcement, caturano2021discovering, chowdary2020autonomous, zennaro2020modeling, ghanem2020reinforcement}. The results of these efforts suggest that reinforcement learning can support the human in charge of the penetration testing process \citep{ghanem2018reinforcement, caturano2021discovering}. Reinforcement learning has also been applied to planning the steps following the initial penetration by learning a policy in a simulated version of the environment \cite{chowdary2020autonomous}. Penetration testing \cite{zennaro2020modeling} and web hacking \cite{erdodi2020agent} have also been converted to simulated capture-the-flag challenges that can be solved with reinforcement learning. Finally, reinforcement learning has been applied to attacking static Portable Executable (PE) \cite{anderson2018learning} and even supervised learning-based anti-malware engines \cite{fang2019evading}. The trained agents are capable of modifying the malware to evade detection. 

Reinforcement learning has also been applied to blue teaming. Microsoft has developed a research toolkit called CyberBattleSim, which enables modeling the behavior of autonomous agents in a high-level abstraction of a computer network \cite{microsoft2021rl}. Reinforcement learning agents that operate in the abstracted network can be trained using the framework. The objective of the platform is to create an understanding of how malicious reinforcement learning agents could behave in a network and how reinforcement learning can be used for threat detection. Deep reinforcement learning can also be applied to improving feature selection for malware detection \cite{fang2019feature}.

Non-learning-based approaches to automating adversary emulation have been developed as well. Caldera is a framework capable of automatically planning adversarial actions against Windows enterprise networks. Caldera uses an inbuilt model of the structure of enterprise domains and knowledge about the objectives and potential actions of an attacker. Then, an intelligent heuristics-based planner decides which adversarial actions to perform \cite{caldera}. Moreover, several different non-RL-based AI approaches to penetration testing and vulnerability analysis have been proposed \cite{mckinnel2019systematic}.

Supervised and unsupervised learning with and without neural networks have been applied for blue teaming purposes. For instance, malicious PowerShell commands can be detected with novel deep learning methods \cite{hendler2018detecting}. An ensemble detector combining an NLP-based classifier with a CNN-based classifier was the best at detecting malicious commands. The detection performance was high enough to be useful in practice. The detector was evaluated using a large dataset consisting of legitimate commands executed by standard users, malicious commands executed by malware, and malicious commands designed by security experts. The suitability of machine learning for intrusion detection, malicious code detection, malware analysis, and spam detection has been discussed \citep{apruzzese2018effectiveness, cui2018detection, kim2018multimodal}. These methods often rely on extensive feature engineering \cite{kim2018multimodal, ccavucsouglu2019new}. In defensive tasks, good performance can often be achieved without deep neural networks, with solutions like logistic regression, support vector machines, and random forests \cite{milosevic2017machine}. Machine learning-based systems are vulnerable to adversarial attacks \cite{chen2019adversarial}, and as different ML-based techniques have dissimilar weaknesses, a combination of machine learning techniques is often necessary \citep{apruzzese2018effectiveness, ccavucsouglu2019new}.

\section{Reinforcement learning}
\label{sec:rl}

Reinforcement learning is one of the three main paradigms of machine learning alongside supervised and unsupervised learning \cite{sutton2018reinforcement}.
In reinforcement learning, an agent interacts with an environment over discrete time steps to maximize its long-run reward. At a given time step $t$, the environment has state $s^{e}_t$ and the agent is given an observation $o_t = f(s^e_t)$ and a reward signal $r_t$. If the environment is fully observable, the observation is equal to the environment state, $o_t = s^e_t$. In a more general scenario, the agent receives only a partial observation $o_t$ which does not represent the full environment state. In this case, the agent has its own state $s_t$ which might differ from the environment state $s^e_t$. The agent selects an action $a_t$ from the set of possible actions $A$ and acts in the environment. The environment transitions to a new state $s^e_{t+1}$ and the agent receives a new observation $o_{t+1}$ and a new reward $r_{t+1}$.
The goal of the agent is to maximize the sum of the collected rewards
$
    R = \sum_{t=0}^{\infty} \gamma^{t} r_{t}
$
where $\gamma \in (0, 1]$ is a discount factor used to discount future rewards \cite{mnih2016asynchronous}.

In this paper, we use a model-free approach to reinforcement learning in which the agent does not build an explicit model of the environment.
The agent selects an action $a_t$ according to a policy function $\pi(a_t | s_t)$ which depends on the agent state $s_t$.
We use an algorithm called the advantage actor-critic (A2C)  \cite{mnih2016asynchronous} in which the policy is parameterized as $\pi(a|s;\theta)$.
The parameters  $\theta$ of the policy are updated in the direction of
\begin{equation}
    \label{eq:grad}
    \nabla_{\theta} \log \pi(a_t | s_t ; \theta)A(a_t, s_t)
\end{equation}
where $A(a_t, s_t)$ is an advantage function which estimates the (relative) benefit of taking action $a_t$ in state $s_t$ in terms of the expected total reward.
In A2C, the advantage function is computed as
\begin{equation*}
    A(a_t, s_t) = \sum_{i=0}^{k-1} \gamma^i r_{t+i} + \gamma^k V(s_{t+k} ) - V(s_t)
\end{equation*}
where $V(s)$ is the state-value function which estimates the expected total reward when the agent starts at state $s_t$ and follows policy $\pi(a_t | s_t ; \theta)$:
\begin{equation*}
    V(s) = \mathbb{E}\left[\sum_{k=0}^{\infty} \gamma^{k}r_{t+k}
    \:\bigg|\: s_t = s \right]
    \,.
\end{equation*}
We update the parameters $\omega$ of the value function $V(s_t; \omega)$ using Monte Carlo estimates of the total discounted rewards as the targets
\[
  \text{loss}\left(V(s_t; \omega), \sum_{k=0}^{\infty} \gamma^{k}r_{t+k}\right)
  \rightarrow \min_{\omega}
\]
using the Huber loss \cite{huber1992robust}. In practice, most of the parameters $\theta$ and $\omega$ are shared (see Section~\ref{sec:arch} and Figure~\ref{fig:nnarch}).

\section{Privilege escalation as a reinforcement learning task}

\subsection{Problem Definition}

In this work, we focus on automating one particular step often performed by red teaming actors: local privilege escalation in Windows. For our reinforcement learning agent, there will be three possible paths to success:
\begin{itemize}
    \item Add the current user as a local administrator
    \item Obtain administrative credentials
    \item Overwrite a program that is executed with elevated privileges when a user or an administrator logs on
\end{itemize}
The first alternative is hardly how a true red teaming actor would approach the problem as changes in the local administrators of a workstation are easily detectable by any advanced detection and response system. However, if the agent is successful at doing that, it demonstrates that the agent can, with some exceptions, execute arbitrary code with elevated privileges on the victim host. The second alternative is a more realistic alternative for performing local privilege escalation. The third method is arguably inferior to the other two methods as it requires the attacker to wait for the system to be rebooted or some other event to occur that triggers the scheduled task or the AutoRun.

\subsection{Learning Environment}
\label{sec:env}

The learning environment is a simulated Windows~7 environment with a random non-zero number of services, tasks, and AutoRuns. In each training episode, we introduce one vulnerability in the simulated system by selecting randomly from the following 12 alternatives:
\begin{enumerate}
    \item hijackable DLL
        \begin{itemize}
        \item missing DLL
        \item writable DLL
        \end{itemize}
    \item re-configurable service
    \item unquoted service path
    \item modifiable ImagePath in the service registry
    \item writable executable pointed to by a service
    \item missing service binary and a writable service folder
    \item writable binary pointed to by an AutoRun
    \item alwaysInstallElevated bits set to one
    \item credentials of a user with elevated access in the WinLogon registry
    \item credentials of a user with elevated access in an Unattend file
    \item writable binary pointed to by a scheduled task running with elevated privileges
    \item writable Startup folder
\end{enumerate}
To increase the variability of the environment states, we also randomly add services, tasks, and AutoRuns that might initially seem vulnerable to the agent. For instance, a service with one of the service-specific vulnerabilities above but without elevated privileges or a service with a writable parent folder but without an unquoted path can be added. Moreover, standard user credentials might be added to the registry, or a folder on the Windows path might be made writable, among others.

To train an autonomous reinforcement learning agent to perform local privilege escalation, we need to formalize the learning problem, that is, we need to define the reward function $r$, the action space $A$, and the space of the agent states.

Defining the reward function is perhaps the easiest task. We selected the simplest possible reward structure without any reward shaping.
The agent is given a reward $r=1$ for the final action of the episode if the privilege escalation has been performed successfully. Otherwise, a zero reward is given. Based on our experiments, this simple sparse reward signal is sufficient for teaching the agent to perform privilege escalation with as few actions as possible because the reward is progressively discounted as more steps are taken by the agent. We also experimented by giving the agent only half a reward ($+\frac{1}{2}$) for performing privilege escalation by the third, arguably inferior method, but the agent had trouble learning the desired behavior.

The state $s^e_t$ of the environment and its dynamics are determined by the Windows~7 environment (or its simulator) that the agent interacts with. The environment is only partially observable: the observations $o_t$ are the outputs of the commands that the agent executes. Working with such a rich observation space is difficult, and therefore, we have designed a custom procedure that converts the observations into the agent state $s_t$. It is the agent state that is used as the input of the policy $\pi(a_t | s_t)$ and value $V(s_t)$ functions. We also manually designed a set $A$ of high-level actions that the agent needs to choose from. We describe the agent state and the action space in the following sections.

\begin{table}
\caption{General information about the system stored in the agent state}
\label{tab:genvariables}
\begin{tabular}{l}
\toprule
Trinary variables:\\
\midrule
\SG{1} Are there credentials in files?\\
\SG{2} Do the credentials in the files belong to users with elevated\\
 \phantom{aal} privileges?\\
\SG{3} Are there credentials in the registry?\\
\SG{4} Do the credentials in the registry belong to users with\\
 \phantom{aal} elevated privileges?\\
\SG{5} Is there a writable folder on the Windows path?\\
\SG{6} Are the AlwaysInstallElevated bits set?\\
\SG{7} Can the AutoRuns be enumerated using an external\\
 \phantom{aal} PowerShell module?\\
\midrule
Binary variables:\\
\midrule
 Has a malicious executable been created in Kali Linux?\\
 Has a malicious service executable been created in Kali Linux?\\
 Has a malicious DLL been created in Kali Linux?\\
 Has a malicious MSI file been created in Kali Linux\\
 Has a malicious executable been downloaded?\\
 Has a malicious service executable been downloaded?\\
 Has a malicious DLL been downloaded?\\
 Has a malicious MSI file been downloaded?\\
 Does the agent know the list of local users?\\
 Are there users whose privileges need to be checked?\\
 Does the agent know the services running on the OS?\\
 Does the agent know the scheduled tasks running on the OS?\\
 Does the agent know the AutoRuns of the OS?\\
 Has the agent performed a static analysis of the service\\
 \phantom{a} binaries to detect DLLs?\\
 Have the DLLs loaded by the service binaries been searched?\\
 Are there folders whose permissions must be checked?\\
 Are there executables whose permissions must be checked?\\
 Does the agent know the current username?\\
 Does the agent know the Windows path?\\
 Are there base64-credentials to decode?\\
\bottomrule
\end{tabular}
\end{table}

\subsection{State of the Agent}
\label{sec:obs}
We update the state of the agent by keeping the information that is relevant for the task of privilege escalation.
The agent state includes variables that contain general information about the system, information about discovered services, dynamic-link libraries (DLLs), AutoRun registry, and scheduled tasks. 

The general information is represented by the 27 variables listed in Table~\ref{tab:genvariables}. Seven of these variables are trinary (true/unknown/false) and contain information useful for the task of privilege escalation. The remaining 20 variables are binary (true/false), and they also contain information about the previous actions of the agent.
The previous actions are included in the state to make the agent state as close to Markov as possible, which makes the training easier.

\begin{table}
  \caption{Part of the agent state with information about services, DLLs, AutoRuns, and scheduled tasks (all are trinary variables)}
  \label{tab:servvariables}
\begin{tabular}{ll}
\toprule
Service & Is the service running? \\
 & Is the service run with elevated privileges? \\
 & Is the service path unquoted? \\
 & Is there a writable parent folder? \\
 & Is there whitespace in the service path? \\
 & Is the service binary in C:{\textbackslash}Windows? \\
 & Can the service executable be written? \\
 & Can the service be re-configured? \\
 & Can the service registry be modified? \\
 & Does the service load a vulnerable DLL? \\
 & Has the service been exploited? \\
\midrule
DLL & Is the DLL missing? \\
 & Is the DLL writable? \\
 & Has the DLL been replaced with a malicious DLL? \\
\midrule
AutoRun & Is the AutoRun file writable? \\
 & Is the AutoRun file in C:\textbackslash Windows? \\
\midrule
Task & Is the task run with elevated privileges? \\
 & Is the executable writable? \\
 & Is the executable in C:\textbackslash Windows? \\
\bottomrule
\end{tabular}
\end{table}

\begin{table}
  \caption{Examples of auxiliary information
  used to fill the command arguments}
  \label{tab:arginfo}
\begin{tabular}{ll}
\toprule
Service & Name, executable path, user\\
Executable & Path, linked DLLs\\
DLL & Name, executable calling, path\\
AutoRun & Executable path, trigger\\
Task & Name, executable path, trigger, user\\
Credentials & Username, password, plaintext\\
File system & Folders, executables, permissions\\
\bottomrule
\end{tabular}
\end{table}

At the beginning of each training episode, the agent has no knowledge of the services running on the host. The agent has to collect a list of services by taking the action \emph{A31 Get a list of services}. Once a service is detected, it is described by its \emph{name}, \emph{full path}, the owning \emph{user} and the 11 trinary attributes listed in Table~\ref{tab:servvariables}. Each of these attributes can have three possible values: true (+1), unknown (0), and false (-1). Then, the agent needs to perform actions such as \emph{A25 Check service permissions with accesschk64} to fill the values of the unknown attributes.

\begin{table*}
  \caption{Actions}
  \label{tab:actions}
  \begin{tabular}{ll}
    \toprule
\AC{1} Create a malicious executable in Kali Linux
 & \AC{19} Change service registry to point to a malicious executable\\
\AC{2} Create a malicious service executable in Kali Linux
 & \AC{20} Change service registry to add the user to local administrators\\
\AC{3} Compile a custom malicious DLL in Kali Linux
 & \AC{21} Install a malicious MSI file\\
\AC{4} Create a malicious MSI in Kali Linux
 & \AC{22} Search for unattend* sysprep* unattended* files\\
\AC{5} Download a malicious executable in Windows
 & \AC{23} Decode base64 credentials\\
\AC{6} Download a malicious service executable in Windows
 & \AC{24} Test credentials\\
\AC{7} Download a malicious DLL in Windows
 & \AC{25} Check service permissions with accesschk64\\
\AC{8} Download a malicious MSI in Windows
 & \AC{26} Check the ACLs of the service registry with Get-ACL\\
\AC{9} Start an exploited service
 & \AC{27} Check executable permissions with icacls\\
\AC{10} Stop an exploited service
 & \AC{28} Check directory permissions with icacls\\
\AC{11} Overwrite the executable of an autorun
 & \AC{29} Analyze service executables for DLLs\\
\AC{12} Overwrite the executable of a scheduled task
 & \AC{30} Search for DLLs\\
\AC{13} Overwrite a service binary
 & \AC{31} Get a list of services\\
\AC{14} Move a malicious executable so that it is executed by
 & \AC{32} Get a list of AutoRuns\\
 \phantom{a} an unquoted service path
 & \AC{33} Get a list of scheduled tasks\\
\AC{15} Overwrite a DLL
 & \AC{34} Check AlwaysInstallElevated bits\\
\AC{16} Move a malicious DLL to a folder on Windows path
 & \AC{35} Check for passwords in Winlogon registry\\
 \phantom{a} to replace a missing DLL
 & \AC{36} Get a list of local users and administrators\\
\AC{17} Re-configure service to use a malicious executable
 & \AC{37} Get the current user \\
\AC{18} Re-configure service to add the user to local administrators
 & \AC{38} Get the Windows path\\
\bottomrule
\end{tabular}
\end{table*}

Since local privilege escalation can be performed by DLL hijacking, we also include the information about the DLLs used by the services in the state. Each DLL is described using a set of attributes listed in Table~\ref{tab:servvariables}. This information is added to the state after taking action \emph{A29 Analyze service executables for DLLs}.

Privileges can be elevated in Windows by using vulnerable executables in the AutoRun registry and misconfigured scheduled tasks. Therefore, we add information about the AutoRun files and the scheduled tasks to the agent state. Each AutoRun file and each scheduled task is described using the trinary attributes defined in Table~\ref{tab:servvariables}.

In addition to the variables defined in Table~\ref{tab:genvariables} and Table~\ref{tab:servvariables}, the agent maintains a collection of auxiliary information in its memory. The information is needed to fill the arguments of the commands executed by the agent. Examples of the auxiliary information are given in Table~\ref{tab:arginfo}. This information is gathered and updated based on the observations, that is, the outputs of the commands performed by the agent. The auxiliary information is not given as input to the neural network, and hence, it affects neither the policy $\pi(a_t | s_t)$ nor the value $V(s_t)$ directly.

\subsection{Action Space}

We designed the action space of the agent by including 
actions needed for gathering information about the victim Windows host and performing the privilege escalation techniques.
The action space consists of 38 actions listed in Table~\ref{tab:actions}. Although the action space is crafted for known privilege escalation vulnerabilities (which we consider unavoidable within the constraints of the current RL), there is no one-to-one relationship between the actions and vulnerabilities. Some actions are only relevant for specific vulnerabilities, whereas many others are more general and can be used in multiple scenarios (see Appendix~\ref{app:actions}). Our general principle in constructing the action space has been to make the actions as atomic as possible while keeping the problem potentially solvable by the current RL. 

The actions are defined on a high level, which means that their exact implementation can vary, for example, depending on the platform. For instance, action \emph{A29 Analyze service executables for DLLs} can be implemented by static analysis of the Portable Executable files with an open-source analyzer to detect the loaded DLLs. The same action can be implemented using a custom analyzer or a script to download the executable and analyze it with Process Monitor.
Our high-level action definition enables modifying the low-level implementations of the actions, such as changing the frameworks used, without affecting the trained agent.
To create the necessary malicious executables, we use Kali Linux with Metasploit. The malicious DLLs needed for performing DLL hijacking are compiled manually. However, the low-level implementation of these commands can easily be changed if desired.

Each of the high-level actions is well-specified and can be performed using only a handful of standard Windows (cmd.exe and PowerShell) and Linux (zsh) commands.
Many of the commands need arguments. For instance, to take the action \emph{A9 Start an exploited service}, the name of the service must be specified. In this work, we automatically fill the arguments using the auxiliary information collected as discussed in Section~\ref{sec:obs}. For example, one of the actions defined is \emph{A28 Check directory permissions with icacls}. The agent maintains an internal list of directories that are of interest, and when the action to analyze the permissions of directories is selected, every directory on the list is scanned. 

\section{Experiments}

\subsection{Simulator of a Windows~7 Virtual Machine}
\label{sec:sim}

A key practical challenge for training a reinforcement learning agent to perform red teaming tasks is the slow simulation speed when performing actions on a real virtual machine.
For example, running commands necessary for privilege escalation can take longer than a minute on a full-featured Windows 7 virtual machine, even if the agent acts optimally.
At the beginning of training, when the agent selects actions very close to randomly, one episode of training on a real VM can last significantly longer. 
Moreover, each training episode requires a new virtual machine that has been configured with one of the available vulnerabilities. Provisioning and configuring a virtual machine in such a manner will further add to the time it would take to train the agent.
Training a successful agent may require thousands of training episodes, which can take a prohibitively large amount of time when training on a real operating system.
Developing an infrastructure to tackle the long simulation times on a real system is a significant challenge, and it is left outside the scope of this study.

To alleviate this issue, we implemented a simulated Python environment that emulates the behavior of a genuine Windows~7 operating system relevant to the privilege escalation task. The simulation consists of, among others, a file system with access controls, Windows registry, AutoRuns, scheduled tasks, users, executables, and services. Using this environment, the actions taken by the agent can be simulated in a highly efficient manner. Moreover, creating simulated machines with random vulnerabilities for training requires little programming and computing power and can be done very fast. However, to determine whether training the agent in a simulated environment instead of a real operating system is feasible, the trained agent will be evaluated by testing it on a vulnerable Windows 7 virtual machine.

\tikzset{%
  block/.style    = {draw, thick, rectangle, minimum height = 3em, minimum width = 3em},
  neuron/.style   = {draw, circle, minimum height=2em},
  sum/.style      = {draw, circle, node distance = 1cm}, 
  mlayer/.style   = {rectangle, draw=black,
                     minimum height=.5em, minimum width = 2em, text centered},
}

\begin{figure}[t]
\centering
\definecolor{color1}{HTML}{f7cac9}
\definecolor{color2}{HTML}{92a8d1}
\definecolor{color3}{HTML}{f4e1d2}
\definecolor{color4}{HTML}{ffef96}
\definecolor{color5}{HTML}{d5f4e6}
\definecolor{color6}{HTML}{b9b0b0}
\definecolor{color7}{HTML}{c2d4dd}
\definecolor{color8}{HTML}{b8a9c9}
\definecolor{color9}{HTML}{e06377}

\begin{tikzpicture}[thick, node distance=7mm]
\usetikzlibrary{arrows,calc}

\draw
node at (0,0) (a1) {\small AutoRun$_1$}
node [below of=a1](a2){\small AutoRun$_M$}
node [below of=a2] (s1) {\small Service$_1$}
node [below of=s1, node distance=5mm] (dll11) {\small DLL$_{11}$}
node [below of=dll11, node distance=5mm] (dll12) {\small DLL$_{1n}$}
node [below of=dll12] (s2) {\small Service$_N$}
node [below of=s2, node distance=5mm] (dll21) {\small DLL$_{N1}$}
node [below of=dll21, node distance=5mm] (dll22) {\small DLL$_{Nm}$}
node [below of=dll22] (t1) {\small Task$_1$}
node [below of=t1] (t2) {\small Task$_L$}

node at ($(a1)!0.5!(a2)$) {...}
node at ($(dll11)!0.5!(dll12)$) {...}
node at ($(dll12)!0.5!(s2)$) {...}
node at ($(dll21)!0.5!(dll22)$) {...}
node at ($(t1)!0.5!(t2)$) {...}

node [mlayer, right of=a1, node distance=10mm, fill=color1] (f_a1) {}
node [mlayer, right of=a2, node distance=10mm, fill=color1] (f_a2) {}
node [mlayer, right of=s1, node distance=10mm, fill=color2] (f_s1) {}
node [mlayer, right of=dll11, node distance=10mm, fill=color6] (f_d11) {}
node [mlayer, right of=dll12, node distance=10mm, fill=color6] (f_d12) {}
node [mlayer, right of=s2, node distance=10mm, fill=color2] (f_s2) {}
node [mlayer, right of=dll21, node distance=10mm, fill=color6] (f_d21) {}
node [mlayer, right of=dll22, node distance=10mm, fill=color6] (f_d22) {}
node [mlayer, right of=t1, node distance=10mm, fill=color3] (f_t1) {}
node [mlayer, right of=t2, node distance=10mm, fill=color3] (f_t2) {}

node at ($(f_a1)!0.5!(f_a2) + (1.3,0)$) [mlayer, minimum height=10mm, minimum width=5mm] (max_a){\small max}
node at ($(f_d11)!0.5!(f_d12) + (1.3,0)$) [mlayer, minimum height=10mm, minimum width=5mm] (max_d1){\small max}
node at ($(f_d21)!0.5!(f_d22) + (1.3,0)$) [mlayer, minimum height=10mm, minimum width=5mm] (max_d2){\small max}
node at ($(f_t1)!0.5!(f_t2) + (1.3,0)$) [mlayer, minimum height=10mm, minimum width=5mm] (max_t){\small max}

node [circle, right of=f_s1, node distance=25mm, fill=black, minimum height=1em]  (c1) {}
node [circle, right of=f_s2, node distance=25mm, fill=black, minimum height=1em]  (c2) {}

node [mlayer, right of=c1, node distance=10mm, fill=color4] (p_s2){}
node at ($(c1) + (10mm,7mm)$) [mlayer, fill=color4] (p_s1){}

node [mlayer, right of=c2, node distance=10mm, fill=color5] (v_s1){}
node at ($(c2) + (10mm,-7mm)$) [mlayer, fill=color5] (v_s2){}

node at ($(p_s1)!0.5!(p_s2) + (14mm,0)$) [mlayer, minimum height=12mm, minimum width=12mm] (select){\small select}
node at ($(v_s1)!0.5!(v_s2) + (14mm,0)$) [mlayer, minimum height=12mm, minimum width=12mm] (argmax){}

;

\draw [->,>=stealth] (f_a1.east) -- (f_a1.east-|max_a.west);
\draw [->,>=stealth] (f_a2.east) -- (f_a2.east-|max_a.west);

\draw [->,>=stealth] (f_d11.east) -- (f_d11.east-|max_d1.west);
\draw [->,>=stealth] (f_d12.east) -- (f_d12.east-|max_d1.west);

\draw [->,>=stealth] (f_d21.east) -- (f_d21.east-|max_d2.west);
\draw [->,>=stealth] (f_d22.east) -- (f_d22.east-|max_d2.west);

\draw [->,>=stealth] (f_t1.east) -- (f_t1.east-|max_t.west);
\draw [->,>=stealth] (f_t2.east) -- (f_t2.east-|max_t.west);

\draw [->,>=stealth] (f_s1.east) -- (c1);
\draw [->,>=stealth] (f_s2.east) -- (c2);

\draw [->,>=stealth] (max_a.east) -- (c1);
\draw [->,>=stealth] (max_a.east) -- (c2);

\draw [->,>=stealth] (max_d1.east) -- (c1);
\draw [->,>=stealth] (max_d2.east) -- (c2);

\draw [->,>=stealth] (max_t.east) -- (c1);
\draw [->,>=stealth] (max_t.east) -- (c2);

\draw [->,>=stealth] (c1) -- (p_s1.west);
\draw [->,>=stealth] (c2) -- (p_s2.west);

\draw [->,>=stealth] (c1) -- (v_s1.west);
\draw [->,>=stealth] (c2) -- (v_s2.west);

\draw [->,>=stealth] (p_s1.east) -- (p_s1.east-|select.west);
\draw [->,>=stealth] (p_s2.east) -- (p_s2.east-|select.west);

\draw [->,>=stealth] (v_s1.east) -- (v_s1.east-|argmax.west);
\draw [->,>=stealth] (v_s2.east) -- (v_s2.east-|argmax.west);

\draw [->,>=stealth] (argmax.north) -- (select.south);
\draw [->,>=stealth] (select.east) -- node [above] {$\pi$} ($(select.east) + (0.5, 0)$);
\draw [->,>=stealth] (argmax.east) -- node [above] {$v$} ($(argmax.east) + (0.5, 0)$);
\node at (argmax.north) [anchor=north] {\scriptsize argmax};
\node at (argmax.east) [anchor=east] {\scriptsize max};

\end{tikzpicture}
\caption{The architecture of the A2C agent. Colored boxes represent multilayer perceptrons (same colors denote shared parameters). The black circles represent the concatenation of input signals.}
\Description{The neural networks takes the autoruns, services and tasks as input and outputs the policy and the value}
\label{fig:nnarch}
\end{figure}
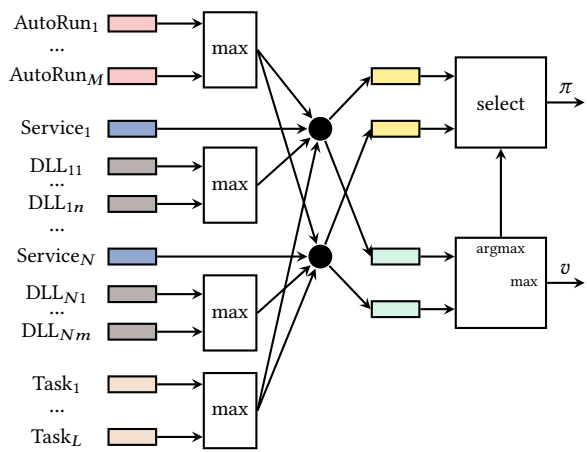

\subsection{The Architecture of the Agent}

\label{sec:arch}

We use an A2C reinforcement learning agent (described in Section~\ref{sec:rl}) in which the policy $\pi(a | s)$ and value $V(s)$ functions are modeled with neural networks. In practice, we use a single neural network with two heads: one produces the estimated value of the state $V(s)$ and the other the probabilities $\pi(a | s)$ of taking one of the 38 actions. The network gets as inputs the state variables described in Tables~\ref{tab:genvariables}~and~\ref{tab:servvariables}. The complete model has less than 27,000 parameters.

The main challenge in designing the neural network is that the number of AutoRuns, services, tasks, and DLLs can vary across training episodes. A Windows host might have anything from dozens to thousands of services, and the number of tasks and AutoRuns might also vary significantly depending on the host. We want our agent to be able to generalize to any number of those.
Therefore, we process the information about each service, AutoRun, and scheduled task separately and aggregate the outputs of these computational blocks using the maximum operation. The architecture of the neural network is presented in Figure~\ref{fig:nnarch}. Computational blocks with shared parameters are shown with the same colors.

In the proposed architecture, we concatenate the max-aggregated outputs of the blocks that process the information about AutoRuns, tasks, and DLLs with the outputs of the blocks that process the information about the individual services. Intuitively, this corresponds to augmenting the service data with the information about the most vulnerable AutoRun and task. Note that we also augment the service data with the information about the DLLs used by the corresponding service. Then, we pass the concatenated information through multilayer perceptrons that output value estimates and policies for all services. Finally, we regard the service with the highest value estimate as the most vulnerable one and select the policy corresponding to that service as the final policy.

\subsection{Training}

Training consists of episodes in which the agent interacts with one instance of a sampled environment. At the beginning of each episode, the agent has no knowledge of the environment. The empty agent state is fed into the neural network that produces the value and the policy outputs. The action is sampled from the probabilities given by the policy output. Thus, we do not use explicit exploration strategies such as epsilon-greedy. The selected action is performed in the simulated environment, and the reward and the observations received as a result of the action are passed to the agent. The observations are parsed to update the agent state as described in Section~\ref{sec:obs}. Then, a new action is selected based on the updated state. The iteration continues until the maximum number of steps for one episode is reached, or the agent has successfully performed privilege escalation.

The parameters of the neural network are updated at the end of each episode.
The gradient \eqref{eq:grad} is computed automatically by backpropagation with PyTorch \cite{NEURIPS2019_9015} and the parameters are updated using the Adam optimizer \cite{kingma2014adam}. The agent is trained for as long as the average reward per episode continues to increase. The hyperparameters used for training the agent are given in Appendix~\ref{app:hyperparameters}.

Figure~\ref{fig:avgeplength} presents the evolution of the 
episode length (averaged over 100 episodes) during one training run.
The average episode length starts from around 200, and it gradually decreases reaching a level slightly above 11 after approximately 30,000 training episodes. We use 1,000 as the maximum number of steps per episode (see Appendix \ref{app:hyperparameters}), which implies that the agent manages to solve the problem and gets rewards from the very beginning of the optimization procedure when it takes close-to-random actions. We estimated that the average episode length is approximately
10.7 actions if the agent acts according to the optimal policy. Thus, the results indicate that the agent has learned to master the task of privilege escalation.

\begin{figure}
\centering
\includegraphics[width=\linewidth]{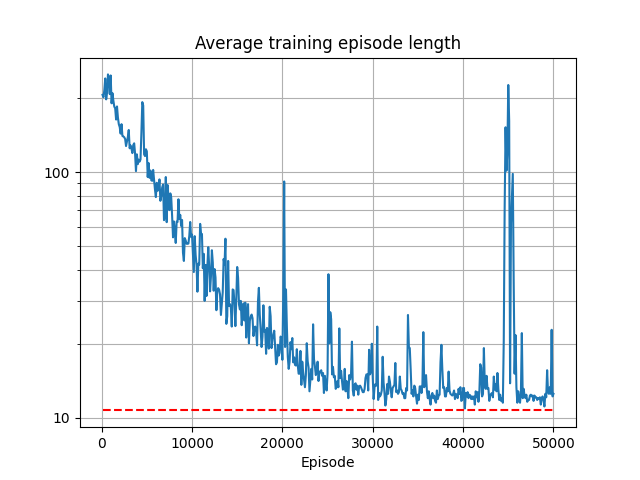}
\caption{The average episode length during training on a logarithmic scale. The red line illustrates the episode length of an optimal policy that is approximately 10.7 actions per episode.}
\Description{The average episode length starts from around 200 and it decreases during the training until it reaches a level slightly above 11 after approximately 30,000 training episodes.}
\label{fig:avgeplength}
\end{figure}

Training the agent for 50,000 episodes in the simulated environment (see Section~\ref{sec:sim}) takes less than two hours without any significant code optimizations using a single NVIDIA GeForce GTX 1080 Ti GPU, which is a high-end consumer GPU from 2017.
Note that performing the same training on a real Windows~7 virtual machine could take weeks. 

\subsection{Testing the Agent}

Next, we test whether the agent trained in our simulated environment can transfer to a real Windows~7 operating system without any adaptation. We also compare the performance of our agent to two baselines: a ruleset crafted by an expert that can be interpreted as the optimal policy and a random policy.

We create a Windows~7 virtual machine using Hyper-V provided by the Windows~10 operating system.
We assume that the offensive actor has gained low-level access with code execution rights by performing, for example, a successful penetration test or a phishing campaign. This is simulated by installing an SSH server on the victim host. We use the Paramiko SSH library in Python to connect to the virtual machine and execute commands with user-level credentials \cite{paramiko}. We use Hyper-V to create a Kali Linux virtual machine with Metasploit for generating malicious executables.
However, instead of using Metasploit for creating malicious DLLs, the agent has to modify and compile a custom DLL code by taking action \emph{A3 Compile a custom malicious DLL in Kali Linux}. Paramiko is also used to connect to the Kali machine.

Using SSH for simulating low-level code execution rights on the victim Windows~7 has some limitations. Some of the Windows command-line utilities such as wmic and accesschk64 are blocked by non-privileged users over SSH. To overcome this limitation of the test scenario, we open a second SSH session using elevated credentials and run the blocked commands in that session. In practice, a malicious actor would be able to execute these utilities while accessing the victim's environment via reverse shell or meterpreter session. Care was taken to prevent the test infrastructure from affecting the target environment. For example, due to the selection of an SSH tunnel 
as an off-the-shelf communication channel for testing purposes, the agent does not target any SSH-related vulnerabilities. Engineering effect was not prioritized for creating a production-ready attack agent, as it was considered beyond the scope of the research.

In order to take some of the actions listed in Table~\ref{tab:actions}, the victim host has to have Windows Sysinternals with accesschk64. Moreover, we need an executable for scanning the DLLs loaded by the PE files. We used an open-source solution for that, but it failed to detect a DLL loaded by a handcrafted service executable. To work around this issue, we hard-coded the result of the scan in the agent. To properly address this issue, the high-level action of performing PE scanning could be mapped to a script that uploads the service executable on a Windows machine and uses ProcMon from Sysinternals to analyze the DLLs loaded by the service executable. Alternatively, a superior PE analyzer could be used.

First, we tested our agent on a virtual machine without external antivirus (AV) software or an intrusion detection system, but which had an up-to-date and active Windows Defender (which is essentially only an anti-spyware program in Windows~7).
We kept the number of services similar to the number of services during training by excluding all services in C:{\textbackslash}Windows from the list of services gathered by the agent. We made the agent deterministically select the action with the highest probability.
Our agent was successful in exploiting all the twelve vulnerabilities. Examples of the sequences of actions taken by the agent during evaluation can be found in Tables~\ref{tab:sequence1}--\ref{tab:sequence4}. 
The agent took very few unnecessary actions.
The performance (measured in terms of the number of actions) could be improved by gathering more information before scanning for directory permissions. Now, the agent prefers scanning the directory permissions immediately after finding interesting directories. However, the amount of noise generated by the agent would have been similar as the agent would have performed more Windows commands per high-level action.

\begin{table}[t]
\caption{Actions to exploit a missing DLL file}
\label{tab:sequence1}
\begin{tabular}{l}
\toprule
\AC{35} Check for passwords in Winlogon registry\\
\AC{37} Get the current user\\
\AC{31} Get a list of services\\
\AC{28} Check directory permissions with icacls\\
\AC{36} Get a list of local users and administrators\\
\AC{26} Check the ACLs of service registries with Get-ACL\\
\AC{25} Check service permissions with accesschk64\\
\AC{34} Check AlwaysInstallElevated bits\\
\AC{32} Get a list of AutoRuns\\
\AC{28} Check directory permissions with icacls\\
\AC{27} Check executable permissions with icacls\\
\AC{22} Search for unattend* sysprep* unattended* files\\
\AC{33} Get a list of scheduled tasks\\
\AC{28} Check directory permissions with icacls\\
\AC{27} Check executable permissions with icacls\\
\AC{29} Analyze service executables for DLLs\\
\AC{30} Search for DLLs\\
\AC{28} Check directory permissions with icacls\\
\AC{38} Get the Windows path\\
\AC{28} Check directory permissions with icacls\\
\AC{3} Compile a custom malicious DLL in Kali Linux\\
\AC{7} Download a malicious DLL in Windows\\
\AC{16} Move a malicious DLL to a folder on Windows path\\
  \phantom{aal} to replace a missing DLL\\
\AC{9} Start an exploited service\\
\bottomrule
\end{tabular}
\end{table}

\begin{table}[t]
\caption{Actions to exploit a service with a missing binary}
\label{tab:sequence2}
\begin{tabular}{l}
\toprule
\AC{35} Check for passwords in Winlogon registry\\
\AC{37} Get the current user\\
\AC{31} Get a list of services\\
\AC{28} Check directory permissions with icacls\\
\AC{2} Create a malicious service executable in Kali Linux\\
\AC{6} Download a malicious service executable in Windows\\
\AC{13} Overwrite a service binary\\
\AC{9} Start an exploited service\\
\bottomrule
\end{tabular}
\end{table}

\begin{table}[t]
\caption{Actions to exploit elevated credentials in the WinLogon registry}
\label{tab:sequence3}
\begin{tabular}{l}
\toprule
\AC{35} Check for passwords in Winlogon registry\\
\AC{36} Get a list of local users and administrators\\
\AC{24} Test credentials\\
\bottomrule
\end{tabular}
\end{table}

\begin{table}[t]
\caption{Actions to exploit AlwaysInstallElevated}
\label{tab:sequence4}
\begin{tabular}{l}
\toprule
\AC{35}  Check for passwords in Winlogon registry\\
\AC{37}  Get the current user\\
\AC{31}  Get a list of services\\
\AC{28}  Check directory permissions with icacls\\
\AC{36}  Get a list of local users and administrators\\
\AC{26}  Check the ACLs of service registries with Get-ACL\\
\AC{25}  Check service permissions with accesschk64\\
\AC{34}  Check AlwaysInstallElevated bits\\
\AC{4}  Create a malicious MSI file in Kali Linux\\
\AC{8}  Download a malicious MSI file in Windows\\
\AC{21}  Install a malicious MSI file\\
\bottomrule
\end{tabular}
\end{table}

After that, we did not limit the number of services (by excluding services in C:{\textbackslash}Windows) and let the agent perform privilege escalation. The increased number of services had no negative effect on the agent, and the agent was successful at the task. An example sequence of commands is given in Appendix~\ref{app:commands}. However, because the agent performs each selected action on every applicable service, the agent generates some noise by scanning through the permissions of all services in C:{\textbackslash}Windows. That could have caused an alert in an advanced detection and response system.

The number of actions used by the agent to escalate privileges during the testing phase is given in Table~\ref{tab:numberOfActions}. We compare the following agents:
\begin{itemize}
\item the oracle agent, which assumes complete knowledge of the system, including the vulnerability;

\item the optimal policy, which is approximated using a fixed ruleset crafted by an expert;

\item the deterministic RL agent, which selects the action with the highest probability;

\item the stochastic RL agent, which samples the action from the probabilities produced by the policy network;

\item an agent taking random actions. 
\end{itemize}
For all random trials, we used 1000 samples and computed the average number of actions used by the agent. Because of the computational cost of running thousands of episodes, all tests involving randomness were run in a simulated environment similar to the testing VM. The results suggest that the policy of the deterministic agent is close to optimal. The addition of stochasticity to action selection has a slightly negative effect on the performance but it increases the variability of the agent's actions making the agent potentially more difficult to detect.

\begin{table*}
\caption{Number of actions used to escalate privileges}
\label{tab:numberOfActions}
\begin{tabular}{c|ccccc}
 Vulnerability & Oracle & Expert & Deterministic & Stochastic & Random \\
  & (full knowledge) & ($\approx$ optimal policy) & RL & RL & \\
 \midrule
 1 & 10 & 20 & 24 & 25.3 & 231.2 \\
 2 & 5 & 10 & 9 & 8.0 & 152.2 \\
 3 & 7 & 7 & 8 & 8.2 & 206.0 \\
 4 & 5 & 9 & 8 & 9.0 & 147.7 \\
 5 & 7 & 10 & 15 & 11.5 & 212.2 \\
 6 & 7 & 7 & 8 & 8.3 & 208.1 \\
 7 & 6 & 9 & 14 & 14.1 & 171.7 \\
 8 & 5 & 15 & 11 & 11.0 & 156.0 \\
 9 & 3 & 11 & 3 & 10.9 & 96.0 \\
 10 & 4 & 13 & 14 & 13.5 & 162.8 \\
 11 & 6 & 9 & 17 & 17.6 & 166.9 \\
 12 & 6 & 8 & 13 & 13.3 & 165.0 \\
 \midrule
 AVG & 5.9 & 10.7 & 12.0 & 12.6 & 173.0 \\
\end{tabular}
\end{table*}

We additionally tested the ability of the agent to generalize to multiple vulnerabilities which might be simultaneously present in the system. This was done in three ways. First, the agent was evaluated in an environment with six different types of vulnerable services present. Second, the agent was evaluated in an environment with all twelve vulnerabilities present. Finally, random combinations of any two vulnerabilities were tested. The agent had little trouble performing privilege escalation in any of these scenarios. 

As a matter of interest, we finally evaluated the agent's performance against a host running an up-to-date version of a standard endpoint protection software, Microsoft Security Essentials, with real-time protection enabled. As expected, the AV software managed to recognize the default malicious executables created by msfvenom in Kali Linux. However, the AV software failed to recognize the custom DLL compiled by the agent, and hence, privilege escalation using DLL hijacking was possible. Moreover, the AV software failed to detect any methods that did not involve a downloaded malicious payload, such as re-configuring the vulnerable service to execute a command that added the user as a local administrator. Hence, privilege escalation was possible in many of the scenarios, even with up-to-date AV software present. It should be noted that these techniques fall beyond the scope of file-based threat detection used by standard antivirus software and would require more advanced protection strategies to counter, such as behavioral- or heuristics-based detection. The agent's performance against such detection engines was considered to be beyond the scope of the project and was not assessed.

\section{Discussion}

Our work demonstrates that it is possible to train a deep reinforcement learning agent to perform local privilege escalation in Windows 7 using known vulnerabilities.
Our method is the first reinforcement learning agent, to the best of our knowledge, that performs privilege escalation with an extensive state space and a high-level action space with easily customizable low-level implementations. Despite being trained in simulated environments, the test results demonstrate that our agent can solve the formalized privilege escalation problem
in a close-to-optimal fashion on full-featured Windows machines with realistic vulnerabilities.

The efficacy of our implementation is limited if up-to-date antivirus software is running on the victim host because only a handcrafted DLL is used, whereas the malicious executables are created using Metasploit with default settings. However, if the mapping from the high-level actions to the low-level commands (see Table \ref{tab:actions}) was improved so that more sophisticated payloads were used or the action space was expanded with actions for defense evasion, a reinforcement learning agent could be capable of privilege escalation in hosts with up-to-date antivirus software but without an advanced detection and response system. 

While simple attacks are likely to be detected by advanced breach detection solutions, not all companies employ those for various reasons. The constant stream of breaches seen in the news reflects that reality. Moreover, if adversaries develop RL-based tools for learning and automating adversarial actions, they might prefer to target networks that are less likely to be running breach detection software.

The current threat level presented by reinforcement learning agents is most likely limited to agents capable of exploiting existing \emph{well-known} vulnerabilities.
The same could be achieved by a scripted agent with a suitable hard-coded logic. However, the RL approach offers a number of benefits compared to a scripted agent:
\begin{itemize}
    \item Scripting an attacking agent can be difficult when the number of potentially exploitable vulnerabilities grows and if the attacked system contains an IDS.
    \item The probabilistic approach of our RL agent will produce more varied attacks (and attempted attacks) than a scripted robot that follows hard-coded rules, which makes our agent more usable for training ML-based defenses and testing and evaluating intrusion detection systems.
    \item An RL agent may be quickly adapted to changes in the environment. For example, if certain sequences of actions cause an alarm raised by an intrusion detection system, the agent might learn to take a different route, which is not detectable by the IDS. This would produce invaluable information for strengthening the defense system. 
\end{itemize}

In the long run, RL agents could have the potential to discover and exploit novel \emph{unseen} vulnerabilities, which would have an enormous impact on the field. To implement this idea, agents would most likely need to interact with an authentic environment, which would require a great deal of engineering effort and a huge amount of computational resources. Crafting the action space would nevertheless be most likely unavoidable within the constraints of the current RL methods. However, the development should go in the direction of making the actions more atomic and minimizing the amount of prior knowledge used in designing the action space. This could allow the agent to encompass more vulnerabilities and could be a way to get closer to the ultimate goal of discovering new vulnerabilities.

Another research direction is to increase the complexity of the learning task. In this first step, we wanted to understand how RL-powered attacks could work in a constrained, varied setup, and our key result is showing that the RL approach works for such a complex learning task. Defeating defensive measures or expanding to a wider range of target environments would be a research topic with a significantly larger scope. It would be interesting to see whether a reinforcement learning agent can perform more steps in the cyber security kill chain, such as defense evasion. It would also be interesting to train the agent in an environment with an intrusion detection system or a defensive RL agent and perform multi-agent reinforcement learning, which has been done in previous research on post-exploitation in simulated environments \cite{elderman2017adversarial}.

\subsection*{Ethical Considerations}

The primary goal of this work is to contribute to building resistant defense systems that can detect and prevent various types of potential attacks.
Rule-based defense systems can be effective, but as the number of attack scenarios grows, they become increasingly difficult to build and maintain. Data-driven defensive systems trained with machine learning offer a promising alternative but implementing this idea in practice is challenged by the problem of scarcity of available training data in this domain. In this work, we present a possible solution to this problem by training a reinforcement learning agent to perform malicious activities and therefore to generate invaluable training data for improving defense systems. The presented approach can also be useful to support the red teaming activities performed by cyber security experts by automating some steps in the kill chain. However, the system developed in this project can potentially be dangerous in the wrong hands. Hence, the code created in this project will not be open-sourced or released to the public.

\begin{acks}
We thank the Security and Software Engineering Research Center S$^2$ERC for funding our work. In addition, we thank David Karpuk, Andrew Patel, Paolo Palumbo, Alexey Kirichenko, and Matti Aksela from F-Secure for their help in running this project and their domain expertise, Tuomas Aura for giving us highly valuable feedback, and the Academy of Finland for the support within the Flagship Programme Finnish Center for Artificial Intelligence (FCAI).
\end{acks}

\bibliographystyle{ACM-Reference-Format}
\bibliography{refs}

\appendix

\section{Supplementary material for RL}

\label{app:hyperparameters}

The hyperparameters used for training are listed below:\\
\begin{tabular}{ccc}
    \toprule
    Parameter&Explanation&Value\\
    \midrule
    $\gamma$ & Discount rate & 0.995 \\
    $\alpha$ & Learning rate for Adam & 0.001 \\
    $\beta_1$ & First moment decay rate for Adam & 0.9 \\
    $\beta_2$ & Second moment decay rate for Adam & 0.999 \\
    $N_{\max}$ & Maximum number of steps & 1000 \\
    $N_{\text{serv}}$ & Number of services & [1, 20] \\
    $N_{\text{auto}}$ & Number of autoruns & [1, 10] \\
    $N_{\text{task}}$ & Number of tasks & [1, 10] \\
    $M_{\text{dlls}}$ & Number of DLLs loaded per service & [1, 4] \\
  \bottomrule
\end{tabular}

\section{Actions Required to Exploit the Vulnerabilities}

\label{app:actions}

The sequences of actions that can be used to exploit the twelve vulnerabilities are presented in this section. The actions used in multiple scenarios are marked with the blue color.
\\
\noindent\textbf{(1.1) Missing DLL}:\\
\begin{tabular}{l}
\midrule
\crow \AC{37} Get the current user  \\
\crow \AC{31} Get a list of services \\
\AC{29} Analyze service executables for DLLs \\
\AC{30} Search for DLLs \\ 
\AC{38} Get the Windows path \\
\crow \AC{28} Check directory permissions with icacls \\
\AC{3} Compile a custom malicious DLL in Kali Linux \\
\AC{7} Download a malicious DLL in Windows \\
\AC{16} Move a malicious DLL to a folder on Windows path\\
  \phantom{a} to replace a missing DLL\\
\crow \AC{9} Start an exploited service \\
\end{tabular}
\\
\noindent\textbf{(1.2) Writable DLL:}\\
\begin{tabular}{l}
\midrule
\crow \AC{37} Get the current user  \\
\crow \AC{31} Get a list of services \\
\AC{29} Analyze service executables for DLLs \\
\AC{30} Search for DLLs \\
\crow \AC{27} Check executable permissions with icacls \\
\AC{3} Compile a custom malicious DLL in Kali Linux \\
\AC{7} Download a malicious DLL in Windows \\
\AC{15} Overwrite a DLL \\
\crow \AC{9} Start an exploited service \\
\end{tabular}
\\
\noindent\textbf{(2) Re-configurable Service:}\\
\begin{tabular}{l}
\midrule
\crow \AC{37} Get the current user \\
\crow \AC{31} Get a list of services \\
\AC{25} Check service permissions with accesschk64 \\
\AC{18} Re-configure service to add the user to local\\
  \phantom{a} administrators \\
\crow \AC{9} Start an exploited service \\
\end{tabular}
\\
\noindent\textbf{(3) Unquoted Service Path:}\\
\begin{tabular}{l}
\midrule
\crow \AC{37} Get the current user \\
\crow \AC{31} Get a list of services \\
\crow \AC{28} Check directory permissions with icacls \\
\crow \AC{2} Create a malicious service executable in Kali Linux \\
\crow \AC{6} Download a malicious service executable in Windows \\
\AC{14} Move a malicious executable so that it is executed by \\
  \phantom{a} an unquoted service path \\
\crow \AC{9} Start an exploited service \\
\end{tabular}
\\
\noindent\textbf{(4) Modifiable ImagePath:}\\
\begin{tabular}{l}
\midrule
\crow \AC{37} Get the current user \\
\crow \AC{31} Get a list of services \\
\AC{26} Check the ACLs of the service registry with Get-ACL \\
\AC{20} Change service registry to add the user to local\\
  \phantom{a} administrators \\
\crow \AC{9} Start an exploited service \\
\end{tabular}
\pagebreak
\\
\noindent\textbf{(5) Writable Service Executable:}\\
\begin{tabular}{l}
\midrule
\crow \AC{37} Get the current user \\
\crow \AC{31} Get a list of services \\
\crow \AC{27} Check executable permissions with icacls \\
\crow \AC{2} Create a malicious service executable in Kali Linux \\
\crow \AC{6} Download a malicious service executable in Windows \\
\crow \AC{13} Overwrite a service binary\\
\crow \AC{9} Start an exploited service \\
\end{tabular}
\\
\noindent\textbf{(6) Missing Service Executable:}\\
\begin{tabular}{l}
\midrule
\crow \AC{37} Get the current user \\
\crow \AC{31} Get a list of services \\
\crow \AC{28} Check directory permissions with icacls \\
\crow \AC{2} Create a malicious service executable in Kali Linux \\
\crow \AC{6} Download a malicious service executable in Windows \\
\crow \AC{13} Overwrite a service binary \\
\crow \AC{9} Start an exploited service \\
\end{tabular}
\\
\noindent\textbf{(7) Writable AutoRun Executable:}\\
\begin{tabular}{l}
\midrule
\crow \AC{37} Get the current user \\
\crow \AC{32} Get a list of AutoRuns \\
\crow \AC{27} Check executable permissions with icacls \\
\crow \AC{1} Create a malicious executable in Kali Linux \\
\crow \AC{5} Download a malicious executable in Windows \\
\crow \AC{11} Overwrite the executable of an AutoRun \\
\end{tabular}
\\
\noindent\textbf{(8) AlwaysInstallElevated:}\\
\begin{tabular}{l}
\midrule
\crow \AC{37} Get the current user \\
\AC{34} Check AlwaysInstallElevated bits \\
\AC{4} Create a malicious MSI in Kali Linux \\
\AC{8} Download a malicious MSI in Windows \\
\AC{21} Install a malicious MSI file \\
\end{tabular}
\\
\noindent\textbf{(9) WinLogon Registry:}\\
\begin{tabular}{l}
\midrule
\AC{35} Check for passwords in Winlogon registry \\
\crow \AC{36} Get a list of local users and administrators \\
\crow \AC{24} Test credentials \\
\end{tabular}
\\
\noindent\textbf{(10) Unattend File:}\\
\begin{tabular}{l}
\midrule
\AC{22} Search for unattend* sysprep* unattended* files \\
\AC{23} Decode base64 credentials \\
\crow \AC{36} Get a list of local users and administrators \\
\crow \AC{24} Test credentials \\
\end{tabular}
\\
\noindent\textbf{(11) Writable Task Binary:}\\
\begin{tabular}{l}
\midrule
\crow \AC{37} Get the current user \\
\AC{33} Get a list of scheduled tasks \\
\crow \AC{27} Check executable permissions with icacls \\
\crow \AC{1} Create a malicious executable in Kali Linux \\
\crow  \AC{5} Download a malicious executable in Windows \\
\AC{12} Overwrite the executable of a scheduled task \\
\end{tabular}
\\
\noindent\textbf{(12) Writable Startup Folder:}\\
\begin{tabular}{l}
\midrule
\crow \AC{37} Get the current user \\
\crow \AC{32} Get a list of AutoRuns \\
\crow \AC{28} Check directory permissions with icacls \\
\crow \AC{1} Create a malicious executable in Kali Linux \\
\crow \AC{5} Download a malicious executable in Windows \\
\crow \AC{11} Overwrite the executable of an AutoRun \\
\end{tabular}

\section{Command-line example}
\label{app:commands}
\makeatletter
\newcommand\footnoteref[1]{\protected@xdef\@thefnmark{\ref{#1}}\@footnotemark}
\makeatother

We exemplify our mapping from actions to commands by showing the commands taken by the agent to escalate privileges by exploiting a service with weak folder permissions and a missing binary (see Table~\ref{tab:sequence2}). 
Note that all the commands disclosed below have been derived from public sources (given in the footnotes) and can be recreated by security practitioners. Furthermore, none of the commands are proprietary to F-Secure.\\
\\
\AC{35} Check for passwords in Winlogon registry:\footnote{https://github.com/sagishahar/lpeworkshop/blob/master/Lab\%20Exercises\%20Walkthrough\%20-\%20Windows.pdf} \\
reg query "HKLM\textbackslash SOFTWARE\textbackslash Microsoft\textbackslash Windows NT\textbackslash \\
\indent CurrentVersion\textbackslash Winlogon" /v DefaultUsername \\
reg query "HKLM\textbackslash SOFTWARE\textbackslash Microsoft\textbackslash Windows NT\textbackslash \\
\indent CurrentVersion\textbackslash Winlogon" /v DefaultPassword
\\
\\
\AC{37} Get the current user:\footnote{\label{fn:note1}https://sushant747.gitbooks.io/total-oscp-guide/content/privilege\_escalation\_windows.html} \\
whoami
\\
\\
\AC{31} Get a list of services:\footnote{https://book.hacktricks.xyz/windows/windows-local-privilege-escalation}\\
wmic service get name,pathname,startname,startmode,started\\
\indent /format:csv
\\
\\
\AC{28} Check directory permissions with icacls:\footnoteref{fn:note1} \\
icacls.exe "c:\textbackslash windows\textbackslash system32" \\
icacls.exe "c:\textbackslash windows" \\
$\dots$ (15 rows skipped)  \\
icacls.exe "c:\textbackslash program files (x86)\textbackslash microsoft\textbackslash edge" \\
icacls.exe "c:\textbackslash program files\textbackslash missing file service" \\
icacls.exe "c:\textbackslash program files" \\
$\dots$ (4 rows skipped) \\
icacls.exe "c:\textbackslash windows\textbackslash system32\textbackslash wbem" \\
icacls.exe "c:\textbackslash program files\textbackslash windows media player"
\\
\\
\AC{2} Create a malicious service executable in Kali Linux:\footnote{\label{fn:note2}https://infosecwriteups.com/privilege-escalation-in-windows-380bee3a2842}  \\
sudo -S msfvenom -p windows/exec CMD='net localgroup \\
\indent administrators user /add' -f exe-service -o java\_updater\_svc \\
\\
\AC{6} Download a malicious service executable in Windows:\footnote{https://adamtheautomator.com/powershell-download-file/}\\
powershell.exe -command "Invoke-WebRequest -Uri \\
\indent '82.130.20.144/java\_updater\_service' \\
\indent -OutFile 'C:\textbackslash Users\textbackslash user\textbackslash Downloads\textbackslash java\_updater\_svc" \\
move /y "C:\textbackslash Users\textbackslash user\textbackslash Downloads\textbackslash java\_updater\_svc"\\
\indent "C:\textbackslash Users\textbackslash user\textbackslash Downloads\textbackslash java\_updater\_svc.exe" \\
\\
\AC{13} Overwrite a service binary:\footnoteref{fn:note2}\\
copy /y "C:\textbackslash Users\textbackslash user\textbackslash Downloads\textbackslash java\_updater\_svc.exe"\\
\indent "c:\textbackslash program files\textbackslash missing file service\textbackslash missingservice.exe"  \\
\\
\AC{9} Start an exploited service:\footnoteref{fn:note2}\\
sc start missingsvc 

\end{document}